\documentclass[reqno,11pt]{amsart}
\usepackage{graphicx}
\usepackage{amscd}
\usepackage[mathscr]{eucal}
\numberwithin{equation}{section}
\allowdisplaybreaks[1]

\newcommand{\SetFigFont}[3]{}

\title[Causal Fermion Systems]{Causal Fermion Systems: A Quantum Space-Time Emerging
from an Action Principle}

\author[F.\ Finster]{Felix Finster}

\author[A.\ Grotz]{Andreas Grotz}

\author[D.\ Schiefeneder]{Daniela Schiefeneder \\ \\ February 2011}
\thanks{Supported in part by the Deutsche Forschungsgemeinschaft.}
\address{Fakult\"at f\"ur Mathematik \\ Universit\"at Regensburg \\ D-93040 Regensburg \\ Germany}
\email{Felix.Finster@mathematik.uni-regensburg.de}
\email{Andreas.Grotz@mathematik.uni-regensburg.de}
\email{Daniela.Schiefeneder@mathematik.uni-regensburg.de}

\newtheorem{Def}{Definition}[section]
\newtheorem{Thm}[Def]{Theorem}

\newtheorem{Lemma}[Def]{Lemma}

\newcommand{\Thanks}{\vspace*{.6em} \noindent \thanks}
\newcommand{\beq}{\begin{equation}}
\newcommand{\eeq}{\end{equation}}
\newcommand{\Proof}{\begin{proof}}
\newcommand{\QED}{\end{proof} \noindent}

\newcommand{\spc}{\;\;\;\;\;\;\;\;\;\;}
\newcommand{\la}{\langle}
\newcommand{\ra}{\rangle}
\newcommand{\bra}{\mathopen{<}}
\newcommand{\ket}{\mathclose{>}}

\newcommand{\C}{\mathbb{C}}
\newcommand{\R}{\mathbb{R}}
\newcommand{\1}{\mbox{\rm 1 \hspace{-1.05 em} 1}}

\newcommand{\N}{\mathbb{N}}

\newcommand{\slsh}{\mbox{ \hspace{-1.13 em} $/$}}

\renewcommand{\H}{\mathscr{H}}
\newcommand{\D}{\mathscr{D}}
\DeclareMathOperator{\Sl}{\prec\!}
\DeclareMathOperator{\Sr}{\!\succ}

\DeclareMathOperator{\norm}{| \hspace*{-0.1em}| \hspace*{-0.1em}|}
\DeclareMathOperator{\Symm}{\mbox{\rm{Symm}}}
\newcommand{\nablaLC}{\nabla^\text{\tiny{\tt{LC}}}}
\newcommand{\DLC}{D^\text{\tiny{\tt{LC}}}}

\renewcommand{\O}{{\mathscr{O}}}
\renewcommand{\L}{{\mathcal{L}}}
\newcommand{\Sact}{{\mathcal{S}}}
\newcommand{\Tact}{{\mathcal{T}}}
\newcommand{\T}{{\mathscr{T}}}

\newcommand{\U}{\text{\rm{U}}}
\newcommand{\Lin}{\text{\rm{L}}}

\newcommand{\F}{{\mathscr{F}}}
\newcommand{\K}{{\mathscr{K}}}

\newcommand{\I}{{\mathcal{I}}}

\begin{document}

\maketitle

\begin{abstract}
Causal fermion systems are introduced as a general mathematical framework for formulating
relativistic quantum theory. By specializing, we recover
earlier notions like fermion systems in discrete space-time, the fermionic projector
and causal variational principles. We review how an effect of spontaneous structure formation
gives rise to a topology and a causal structure in space-time.
Moreover, we outline how to construct a spin connection and curvature, leading to
a proposal for a ``quantum geometry'' in the Lorentzian setting.
We review recent numerical and analytical results on the support of minimizers of causal
variational principles which reveal a ``quantization effect'' resulting in a discreteness of space-time.
A brief survey is given on the correspondence to quantum field theory and gauge theories.
\end{abstract}

\tableofcontents

\section{The General Framework of Causal Fermion Systems}
Causal fermion systems provide a general mathematical framework for the formulation of
relativistic quantum theory. They arise by generalizing and harmonizing earlier notions like the ``fermionic projector,'' ``fermion systems in discrete space-time'' and ``causal variational principles.''
After a brief motivation of the basic objects (Section~\ref{secmot}), we introduce the
general framework, trying to work out the mathematical
essence from an abstract point of view (Sections~\ref{secframe} and~\ref{secaction}).
By specializing, we then recover the earlier notions (Section~\ref{secspecial}).
Our presentation is intended as a mathematical introduction, which can clearly be supplemented by the more
physical introductions in the survey articles~\cite{lrev, dice2010, srev}.

\subsection{Motivation of the Basic Objects} \label{secmot}
In order to put the general objects into a simple and concrete context, we begin with the free
Dirac equation in Minkowski space. Thus we let~$(M, \la .,. \ra)$ be Minkowski space 
(with the signature convention~$(+ - - -)$) and~$d\mu$ the standard volume measure
(thus~$d\mu = d^4x$ in a reference frame~$x= (x^0, \ldots, x^3)$). We consider a
subspace~$I$ of the solution space of the Dirac equation~$(i \gamma^j \partial_j - m) \psi = 0$
($I$ may be finite or infinite dimensional).
On~$I$ we introduce a scalar product~$\la .|. \ra_\H$. The most natural choice is
to take the scalar product associated to the probability integral,
\beq \label{pip}
\la \psi | \phi \ra_\H = 2 \pi \int_{t=\text{const}} (\overline{\psi} \gamma^0 \phi)(t, \vec{x})\, d\vec{x}
\eeq
(where~$\overline{\psi} = \psi^\dagger \gamma^0$ is the usual adjoint spinor; note that due to
current conservation, the value of the integral is independent of~$t$), but other
choices are also possible. In order not to distract from the main ideas,
in this motivation we disregard technical issues by implicitly
assuming that the scalar product~$\la .|. \ra_\H$ is
well-defined on~$I$ and by ignoring the fact that mappings on~$I$ may be defined only
on a dense subspace (for details on how to make the following consideration rigorous
see~\cite[Section~4]{lqg}).
Forming the completion of~$I$, we obtain a Hilbert space~$(\H, \la .|. \ra_\H)$.

Next, for any~$x \in M$ we introduce the sesquilinear form
\beq \label{bdef}
b \::\: \H \times \H \rightarrow \C \::\: (\psi, \phi) \mapsto -(\overline{\psi} \phi)(x) \:.
\eeq
As the inner product~$\overline{\psi} \phi$ on the Dirac spinors is indefinite of signature~$(2,2)$,
the sesquilinear form~$b$ has signature~$(p,q)$ with~$p,q \leq 2$. Thus we may uniquely represent it
as
\beq \label{brep}
b(\psi, \phi) = \la \psi | F \phi \ra_\H
\eeq
with a self-adjoint operator~$F \in \Lin(\H)$ of finite rank,
which (counting with multiplicities) has at most two positive and at most two negative eigenvalues.
Introducing this operator for every~$x \in M$, we obtain a mapping
\beq \label{Fmapdef}
F \::\: M \rightarrow \F\:,
\eeq
where~$\F \subset \Lin(\H)$ denotes the set of all self-adjoint operators of finite rank with
at most two positive and at most two negative eigenvalues, equipped with the topology induced
by the Banach space~$\Lin(\H)$.

It is convenient to simplify this setting in the following way.
In most physical applications, the mapping~$F$ will be
injective with a closed image. Then we can identify~$M$ with the
subset~$F(M) \subset \F$. Likewise, we can identify the measure~$\mu$
with the push-forward measure~$\rho=F_*\, \mu$ on~$F(M)$
(defined by~$\rho(\Omega) = \mu(F^{-1}(\Omega))$).
The measure~$\rho$ is defined even on all of~$\F$, and the image of~$F$
coincides with the support of~$\rho$.
Thus setting~$M = \text{supp}\, \rho$, we can reconstruct space-time from~$\rho$.
This construction allows us to describe the physical
system by a single object: the measure~$\rho$ on~$\F$.
Moreover, we can extend our notion of space-time simply by allowing~$\rho$ to be a more
general measure (i.e.\ a measure which can no longer be realized as the push-forward~$\rho=F_*\, \mu$
of the volume measure in Minkowski space with a continuous function~$F$).

We have the situation in mind when~$I$ is composed of all the occupied fermionic states
of a physical system, including the states of the Dirac sea (for a physical discussion see~\cite{srev}).
In this situation, the causal structure is encoded in the spectrum of the operator product~$F(x) \!\cdot\! F(y)$.
In the remainder of this section, we explain how this works for the vacuum.
Thus let us assume that~$I$ is the subspace of all negative-energy solutions of the Dirac equation.
We first compute~$F$.
\begin{Lemma} \label{lemmaF}
Let~$\psi, \phi$ be two smooth negative-energy solutions of the free Dirac equation. We set
\[ \big( F(y) \,\phi \big)(x) = P(x,y)\: \phi(y) \:, \]
where~$P(x,y)$ is the distribution
\beq \label{Psea}
P(x,y) = \int \frac{d^4k}{(2 \pi)^4}\: (k \slsh+m)\:
\delta(k^2-m^2)\: \Theta(-k^0)\: e^{-ik(x-y)} \:.
\eeq
Then the equation
\[ \la \psi | F(y) \,\phi \ra_\H = -(\overline{\psi} \phi)(y) \]
holds, where all integrals are to be understood in the distributional sense.
\end{Lemma}
\Proof We can clearly assume that~$\psi$ is a plane-wave solution, which
for convenience we write as
\beq \label{pplane}
\psi(x) = (q\slsh+m)\: \chi\: e^{-i q x} \:,
\eeq
where~$\chi$ is a constant spinor. Here~$q=(q^0, \vec{q})$
with~$\vec{q} \in \R^3$ and $q^0 = -\sqrt{|\vec{q}|^2 + m^2}$
is a momentum on the lower mass shell. A straightforward calculation yields
\begin{align*}
\la \psi &| F(y) \,\phi \ra_\H \overset{\eqref{pip}}{=} 2 \pi
\int_{\R^3} d\vec{x} \;\overline{\chi}\, (q\slsh+m) \: e^{i q x} \:\gamma^0\, P(x,y)\, \phi(y) \\
&\overset{\eqref{Psea}}{=} 
\int d^4k \: \delta^3(\vec{k} - \vec{q})\:
\overline{\chi}\, (q\slsh+m) \gamma^0 (k \slsh+m)\: \delta(k^2-m^2)\: \Theta(-k^0)\: e^{ik y} \phi(y) \\
&\:= \frac{1}{2 |q^0|}\: \overline{\chi}\, (q\slsh+m) \gamma^0 (q \slsh+m)\: e^{iq y} \phi(y) \\
&\overset{(\ast)}{=} -\overline{\chi} \,(q\slsh+m)\: e^{iq y} \,\phi(y) = -(\overline{\psi} \phi)(y)\:,
\end{align*}
where in~($\ast$) we used the anti-commutation relations of the Dirac matrices.
\QED
This lemma gives an explicit solution to~\eqref{brep} and~\eqref{bdef}. The fact that~$F(y) \phi$
is merely a distribution shows that an ultraviolet regularization is needed in order for~$F(y)$
to be a well-defined operator on~$\H$.
We will come back to this technical point after~\eqref{Pseaeps} and refer to~\cite[Section~4]{lqg}
for details. For clarity, we now proceed simply by computing the eigenvalues of
the operator product~$F(y)\!\cdot\! F(x)$
formally (indeed, the following calculation is mathematically rigorous except if~$y$ lies on the
boundary of the light cone centered at~$x$, in which case the expressions become singular).
First of all, as the operators~$F(y)$ and~$F(x)$ have rank at most four,
we know that their product~$F(y) \!\cdot\! F(x)$ also has at most four non-trivial eigenvalues,
which counting with algebraic multiplicities we denote by~$\lambda_1 \ldots, \lambda_4$.
Since this operator product is self-adjoint only if the factors~$F(x)$ and~$F(y)$ commute, the
eigenvalues~$\lambda_1, \ldots, \lambda_4$ will in general be complex.
By iterating Lemma~\ref{lemmaF}, we find that for any~$n \geq 0$,
\[ \Big( F(x)\, \big(F(y)\, F(x) \big)^n \phi \Big)(z) = P(z,x)\: \Big( P(x,y)\, P(y,x) \Big)^n \phi(x) \:. \]
Forming a characteristic polynomial, one sees
that the non-trivial eigenvalues of $F(y) \!\cdot\! F(x)$ coincide precisely with the
eigenvalues of the $(4 \times 4)$-matrix~$A_{xy}$ defined by
\[ A_{xy} = P(x,y)\, P(y,x)\:. \]
The qualitative properties of the eigenvalues of~$A_{xy}$ can easily be determined
even without computing the Fourier integral~\eqref{Psea}:
From Lorentz symmetry, we know that for all~$x$ and~$y$ for which the Fourier integral
exists, $P(x,y)$ can be written as
\beq \label{Pxyrep}
P(x,y) \;=\; \alpha\, (y-x)_j \gamma^j + \beta\:\1
\eeq
with two complex coefficients~$\alpha$ and~$\beta$. Taking the conjugate, we see that
\[ P(y,x) \;=\; \overline{\alpha}\, (y-x)_j \gamma^j + \overline{\beta}\:\1 \:. \]
As a consequence,
\beq \label{1}
A_{xy} \;=\; P(x,y)\, P(y,x) \;=\; a\, (y-x)_j \gamma^j + b\, \1
\eeq
with two real parameters~$a$ and $b$ given by
\beq \label{ab}
a \;=\; \alpha \overline{\beta} + \beta \overline{\alpha} \:,\spc
b \;=\; |\alpha|^2 \,(y-x)^2 + |\beta|^2 \:.
\eeq
Applying the formula~$(A_{xy} - b \1)^2 = a^2\:(y-x)^2\,\1$,  we find that
the roots of the characteristic polynomial of~$A_{xy}$ are given by
\[ b \pm \sqrt{a^2\: (y-x)^2} \:. \]
Thus if the vector~$(y-x)$ is timelike, the term~$(y-x)^2$ is positive, so that
the~$\lambda_j$ are all real. If conversely the vector~$(y-x)$ is spacelike,
the term~$(y-x)^2$ is negative, and the~$\lambda_j$ form a complex conjugate
pair. We conclude that the the causal structure of Minkowski space
has the following spectral correspondence:
\beq
\begin{split}
&\text{The non-trivial eigenvalues of~$F(x) \!\cdot\! F(y)$} \\
&\!\left\{\!\! \begin{array}{c}
\text{are real} \\ \text{form a complex conjugate pair} \end{array} \!\!\right\}\!
\text{ if $x$ and~$y$ are }
\!\left\{ \!\!\begin{array}{c}
\text{timelike} \\
\text{spacelike}
\end{array} \!\!\right\}\! \text{ separated}.
\end{split}
\eeq

\subsection{Causal Fermion Systems} \label{secframe}
Causal fermion systems have two formulations, referred to as the particle
and the space-time representation. We now introduce both formulations and explain their relation.
After that, we introduce the setting of the fermionic projector as a special case.

\subsubsection{From the Particle to the Space-Time Representation}
\begin{Def} \label{defparticle} {\em{
Given a complex Hilbert space~$(\H, \la .|. \ra_\H)$ (the {\em{``particle space''}})
and a parameter~$n \in \N$ (the {\em{``spin dimension''}}), we let~$\F \subset \Lin(\H)$ be the set of all
self-adjoint operators on~$\H$ of finite rank, which (counting with multiplicities) have
at most~$n$ positive and at most~$n$ negative eigenvalues. On~$\F$ we are given
a positive measure~$\rho$ (defined on a $\sigma$-algebra of subsets of~$\F$), the so-called
{\em{universal measure}}. We refer to~$(\H, \F, \rho)$ as a {\em{causal fermion system in the
particle representation}}.
}}
\end{Def} \noindent
Vectors in the particle space have the interpretation as the occupied fermionic states
of our system. The name ``universal measure'' is motivated by the fact that~$\rho$ describes the
distribution of the fermions in a space-time ``universe'', with causal relations defined as
follows.
\begin{Def} (causal structure) \label{def2}
{\em{ For any~$x, y \in \F$, the product~$x y$ is an operator
of rank at most~$2n$. We denote its non-trivial eigenvalues (counting with algebraic multiplicities)
by~$\lambda^{xy}_1, \ldots, \lambda^{xy}_{2n}$. The points~$x$ and~$y$ are
called {\em{timelike}} separated if the~$\lambda^{xy}_j$ are all real. They are said to be
{\em{spacelike}} separated if all the~$\lambda^{xy}_j$ are complex
and have the same absolute value.
In all other cases, the points~$x$ and~$y$ are said to be {\em{lightlike}} separated. }}
\end{Def} \noindent
Since the operators~$xy$ and~$yx$ are isospectral (this follows from the matrix
identity~$\det(BC-\lambda \1)=\det(CB-\lambda \1)$; see for
example~\cite[Section~3]{discrete}), this definition is symmetric in~$x$ and~$y$.

We now construct additional objects, leading us to the more familiar space-time
representation. First, on~$\F$ we consider the topology induced by the
operator norm~$\|A\| := \sup \{ \|A u \|_\H \text{ with } \| u \|_\H = 1 \}$.
For every~$x \in \F$
we define the {\em{spin space}}~$S_x$ by~$S_x = x(\H)$; it is a subspace of~$\H$ of dimension
at most~$2n$. On~$S_x$ we introduce the {\em{spin scalar product}} $\Sl .|. \Sr_x$ by
\beq \label{ssp}
\Sl u | v \Sr_x = -\la u | x u \ra_\H \qquad \text{(for all $u,v \in S_x$)}\:;
\eeq
it is an indefinite inner product of signature~$(p,q)$ with~$p,q \leq n$. We define
{\em{space-time}}~$M$ as the support of the universal measure, $M = \text{supp}\, \rho$.
It is a closed subset of~$\F$, and by restricting the causal structure of~$\F$ to~$M$, we
get causal relations in space-time. A {\em{wave function}}~$\psi$ is defined as a function
which to every~$x \in M$ associates a vector of the corresponding spin space,
\beq \label{psirep}
\psi \::\: M \rightarrow \H \qquad \text{with} \qquad \psi(x) \in S_x \quad \text{for all~$x \in M$}\:. 
\eeq
On the wave functions we introduce the indefinite inner product
\beq \label{Sprod}
\bra \psi | \phi \ket = \int_M \Sl \psi(x) | \phi(x) \Sr_x \: d\rho(x) \:.
\eeq
In order to ensure that the last integral converges, we also introduce the norm~$\norm . \norm$  by
\beq \label{ndef}
\norm \psi \norm^2 = \int_M \la \psi(x) |\, |x|\, \psi(x) \ra_\H \:d\rho(x)
\eeq
(where~$|x|$ is the absolute value of the operator~$x$ on~$\H$).
The {\em{one-particle space}} $\K$ is defined as the space of wave functions for which
the norm~$\norm . \norm$ is finite, with the topology induced by this norm, and endowed with
the inner product~$\bra .|. \ket$. Then~$(\K, \bra .|. \ket)$ is a Krein space (see~\cite{bognar}).
Next, for any~$x, y \in M$ we define the kernel of the fermionic operator~$P(x,y)$ by
\beq \label{Pxydef}
P(x,y) = \pi_x \,y|_{S_y} \::\: S_y \rightarrow S_x \:,
\eeq
where~$\pi_x$ is the orthogonal projection onto the subspace~$S_x \subset \H$
(and~$|_{S_y}$ denotes the restriction of an operator to~$S_y$).
The {\em{closed chain}} is defined as the product
\[ A_{xy} = P(x,y)\, P(y,x) \::\: S_x \rightarrow S_x\:. \]
As it is an endomorphism of~$S_x$, we can compute its eigenvalues.
The calculation~$A_{xy} = (\pi_x y)(\pi_y x) = \pi_x\, yx$ shows that these eigenvalues
coincide precisely with the non-trivial eigenvalues~$\lambda^{xy}_1, \ldots, \lambda^{xy}_{2n}$
of the operator~$xy$ as considered in Definition~\ref{def2}. In this way, the kernel of the fermionic
operator encodes the causal structure of~$M$.
Choosing a suitable dense domain of definition\footnote{For example, one may choose~$\D(P)$
as the set of all vectors~$\psi \in \K$ satisfying the conditions
\[ \phi := \int_M x\, \psi(x)\, d\rho(x) \:\in \: \H \qquad \text{and} \qquad \norm \phi \norm < \infty\:. \]
}~$\D(P)$, we can regard~$P(x,y)$ as the integral kernel
of a corresponding operator~$P$,
\beq \label{Pdef}
P \::\: \D(P) \subset \K \rightarrow \K \:,\qquad (P \psi)(x) =
\int_M P(x,y)\, \psi(y)\, d\rho(y)\:,
\eeq
referred to as the {\em{fermionic operator}}. We collect two properties of the fermio\-nic operator:
\begin{itemize}
\item[(A)] $P$ is {\em{symmetric}} in the sense that~$\bra P \psi | \phi \ket = \bra \psi | P \phi \ket$
for all~$\psi, \phi \in \D(P)$: \\ \label{ABdef}
According to the definitions~\eqref{Pxydef} and~\eqref{ssp},
\begin{align*}
\Sl P(x,y) \,\psi(y) \,|\, \psi(x) \Sr_x &= - \la (\pi_x \,y \,\psi(y)) \,|\, x \,\phi(x) \ra_\H \\
&= - \la \psi(y) \,|\, y x \,\phi(x) \ra_\H = \Sl \psi(y) \,|\,  P(y,x) \,\psi(x) \Sr_y\:.
\end{align*}
We now integrate over~$x$ and~$y$ and apply~\eqref{Pdef} and~\eqref{Sprod}.
\item[(B)] $(-P)$ is {\em{positive}} in the sense that~$\bra \psi | (-P) \psi \ket \geq 0$~for
all $\psi \in \D(P)$: \\
This follows immediately from the calculation
\begin{align*}
\bra \psi | (-P) \psi \ket &= - \iint_{M \times M} \Sl \psi(x) \,|\, P(x,y)\, \psi(y) \Sr_x\:
d\rho(x)\, d\rho(y) \\
&= \iint_{M \times M} \la \psi(x) \,|\, x \, \pi_x \,y\, \psi(y) \ra_\H \:
d\rho(x)\, d\rho(y) = \la \phi | \phi \ra_\H \geq 0 \:,
\end{align*}
where we again used~\eqref{Sprod} and~\eqref{Pxydef} and set
\[ \phi = \int_M x\, \psi(x)\: d\rho(x)\:. \]
\end{itemize}
The {\em{space-time representation}} of the causal fermion system consists of
the Krein space~$(\K, \bra .|. \ket)$, whose vectors are represented as functions
on~$M$ (see~\eqref{psirep}, \eqref{Sprod}), together with the fermionic operator~$P$
in the integral representation~\eqref{Pdef} with the above properties~(A) and~(B).

Before going on, it is instructive to consider the symmetries of our framework.
First of all, unitary transformations
\[ \psi \rightarrow U \psi \qquad \text{with~$U \in \Lin(\H)$ unitary} \]
give rise to isomorphic systems. This symmetry corresponds to the fact that the fermions are
{\em{indistinguishable particles}} (for details see~\cite[\S3.2]{PFP} and~\cite[Section~3]{entangle}).
Another symmetry becomes apparent if we choose
basis representations of the spin spaces and write the wave functions in components.
Denoting the signature of~$(S_x, \Sl .|. \Sr_x)$ by~$(p(x),q(x))$, we choose
a pseudo-orthonormal basis~$(\mathfrak{e}_\alpha(x))_{\alpha=1,\ldots, p+q}$ of~$S_x$,
\[ \Sl \mathfrak{e}_\alpha | \mathfrak{e}_\beta \Sr = s_\alpha\: \delta_{\alpha \beta}
\qquad \text{with} \qquad
s_1, \ldots, s_p = 1\:,\;\;\; s_{p+1}, \ldots, s_{p+q}=-1 \:. \]
Then a wave function~$\psi \in \K$ can be represented as
\[ \psi(x) = \sum_{\alpha=1}^{p+q} \psi^\alpha(x)\: \mathfrak{e}_\alpha(x) \]
with component functions~$\psi^1, \ldots, \psi^{p+q}$.
The freedom in choosing the basis~$(\mathfrak{e}_\alpha)$ is described by the
group~$\U(p,q)$ of unitary transformations with respect to an inner product of signature~$(p,q)$,
\beq \label{lgf}
\mathfrak{e}_\alpha \rightarrow \sum_{\beta=1}^{p+q} (U^{-1})^\beta_\alpha\;
\mathfrak{e}_\beta \qquad \text{with $U \in \U(p,q)$}\:.
\eeq
As the basis~$(\mathfrak{e}_\alpha)$ can be chosen independently at each space-time point,
this gives rise to local unitary transformations of the wave functions,
\beq \label{lgt}
\psi^\alpha(x) \rightarrow  \sum_{\beta=1}^{p+q} U(x)^\alpha_\beta\: \psi^\beta(x)\:.
\eeq
These transformations can be interpreted as {\em{local gauge transformations}}
(see also Section~\ref{secQFT}). Thus in our framework, the gauge group is
the isometry group of the spin scalar product; it is a non-compact group
whenever the spin scalar product is indefinite. {\em{Gauge invariance}} is incorporated
in our framework simply because the basic definitions are basis independent.

The fact that we have a distinguished representation of the wave functions as functions on~$M$
can be expressed by the {\em{space-time projectors}}, defined as the operators of multiplication
by a characteristic function. Thus for any measurable~$\Omega \subset M$, we define
the space-time projector~$E_\Omega$ by
\[ E_\Omega \::\: \K \rightarrow \K\:, \qquad (E_\Omega \psi)(x) = \chi_\Omega(x)\, \psi(x)\:. \]
Obviously, the space-time projectors satisfy the relations
\beq \label{Erel}
E_U E_V = E_{U \cap V}\:,\quad E_U + E_V = E_{U \cup V} + E_{U \cap V} \:,\qquad
E_M = \1_\K \:,
\eeq
which are familiar in functional analysis as the relations which characterize spectral projectors.
We can now take the measure space~$(M, \rho)$
and the Krein space~$(\K, \bra .|.\ket)$ together with the fermionic operator and the
space-time projectors as the abstract starting point.

\subsubsection{From the Space-Time to the Particle Representation}
\begin{Def} \label{def14}
{\em{ Let~$(M, \rho)$ be a measure space ({\em{``space-time''}}) and~$(\K, \bra .|. \ket)$ 
a Krein space (the {\em{``one-particle space''}}). Furthermore, we let~$P : \D(P) \subset \K \rightarrow \K$ 
be an operator with dense domain of definition~$\D(P)$ (the {\em{``fermionic operator''}}),
such that~$P$ is symmetric and~$(-P)$ is positive (see~(A) and~(B) on page~\pageref{ABdef}).
Moreover, to every $\rho$-measurable set~$\Omega \subset M$ we associate a projector~$E_\Omega$
onto a closed subspace~$E_\Omega(\K) \subset \K$, such that the 
resulting family of operators~$(E_\Omega)$ (the {\em{``space-time projectors''}}) satisfies
the relations~\eqref{Erel}.
We refer to~$(M, \rho)$ together with~$(\K, \bra .|. \ket, E_\Omega, P)$ as a {\em{causal fermion
system in the space-time representation}}.
}} \end{Def}

This definition is more general than the previous setting because it does not involve a notion of
spin dimension. Before one can introduce this notion, one needs to ``localize'' the
vectors in~$\K$ with the help of the space-time projectors to obtain wave functions on~$M$.
If~$\rho$ were a discrete measure, this localization could be obtained by considering the
vectors~$E_x \psi$ with~$x \in \text{supp}\, \rho$. If~$\psi$ could be expected to be a continuous
function, we could consider the vectors~$E_{\Omega_n} \psi$ for~$\Omega_n$ a decreasing sequence of
neighborhoods of a single point. In the general setting of Definition~\ref{def14}, however,
we must use a functional analytic construction, which in the easier Hilbert space setting
was worked out in~\cite{gauge}. We now sketch how the essential
parts of the construction can be carried over to Krein spaces.
First, we need some technical assumptions.
\begin{Def} \label{deftechnical}
{\em{ A causal fermion system in the space-time representation has {\em{spin dimension
at most~$n$}} if there are vectors~$\psi_1, \ldots, \psi_{2n} \in \K$ with the following properties:
\begin{itemize}
\item[(i)] For every measurable set~$\Omega$, the matrix~$S$ with 
components~$S_{ij} = \bra \psi_i | E_\Omega \psi_j \ket$ has at most~$n$ positive and at most~$n$
negative eigenvalues.
\item[(ii)] The set
\beq \label{Esets}
\left\{ E_\Omega \psi_k \text{ with~$\Omega$ measurable
and~$k=1,\ldots, 2n$} \right\}
\eeq
generates a dense subset of~$\K$.
\item[(iii)] For all~$j, k \in \{1, \ldots, 2n\}$, the mapping
\[ \mu_{jk} :  \Omega \rightarrow \bra \psi_j | E_\Omega \psi_k \ket \]
defines a complex measure on~$M$ which is absolutely continuous with respect to~$\rho$.
\end{itemize}
}} \end{Def} \noindent
This definition allows us to use the following construction. In order to introduce the spin
scalar product between the vectors~$\psi_1, \ldots \psi_{2n}$, we use property~(iii)
to form the Radon-Nikodym decomposition
\[ \bra \psi_j | E_\Omega \,\psi_k \ket = \int_\Omega \Sl \psi_j | \psi_k \Sr_x\: d\rho(x) 
\quad \text{with} \quad \Sl \psi_j | \psi_k \Sr \in L^1(M, d\rho)\:, \]
valid for any measurable set~$\Omega \subset M$. Setting
\[ \Sl E_U \psi_j | E_V \psi_k \Sr_x = \chi_U(x)\: \chi_V(x)\; \Sl \psi_j | \psi_k \Sr_x \:, \]
we can extend the spin scalar product to the sets~\eqref{Esets}. Property~(ii)
allows us to extend the spin scalar product by approximation to all of~$\K$.
Property~(i) ensures that the spin scalar product has the signature~$(p,q)$ with~$p,q \leq n$.

Having introduced the spin scalar product, we can now get a simple connection to the particle
representation: The range of the fermionic operator~$I:= P(\D(P))$ is a (not necessarily closed) subspace
of~$\K$. By
\[ \la P(\phi) \,|\, P(\phi') \ra := \bra \phi | (-P) \phi' \ket \]
we introduce on~$I$ an inner product~$\la .|. \ra$, which by the positivity property~(B) is positive
semi-definite. Thus its abstract completion~$\H := \overline{I}$ is a Hilbert space~$(\H, \la .|. \ra_\H$).
We again let~$\F \subset \Lin(\H)$ be the set of all self-adjoint operators of finite rank, which have
at most~$n$ positive and at most~$n$ negative eigenvalues. For any~$x \in M$, the conditions
\beq \label{Fdef}
\la \psi | F \phi \ra_\H = -\Sl \psi | \phi \Sr_x \quad \text{for all~$\psi, \phi \in I$}
\eeq
uniquely define a self-adjoint operator~$F$ on~$I$, which has finite rank and at most~$n$ positive
and at most~$n$ negative eigenvalues. By continuity, this operator uniquely extends to an
operator~$F \in \F$, referred to as the {\em{local correlation operator}} at~$x$. We thus obtain
a mapping~$F \::\: M \rightarrow \F$. Identifying points of~$M$ which have the same
image (see the discussion below), we can also consider the subset~$\F(M)$ of~$\F$ as our space-time.
Replacing~$M$ by~$F(M)$ and~$\rho$ by the push-forward measure~$F_* \rho$ on~$\F$,
we get back to the setting of Definition~\ref{defparticle}.

We point out that, despite the fact that the particle and space-time representations can be constructed
from each other, the two representations are {\em{not}} equivalent. The reason is that
the construction of the particle representation involves an identification of points of~$M$
which have the same local correlation operators.
Thus it is possible that two causal fermion systems in the
space-time representation which are not gauge equivalent may have the same particle
representation\footnote{As a simple example consider the case~$M=\{0,1\}$ with~$\rho$ the
counting measure, $\K = \C^4$ with~$\bra\! \psi | \phi \!\ket = \la \psi, S \phi \ra_{\C^4}$ and the signature
matrix~$S=\text{diag}(1,-1,1,-1)$. Moreover, we choose the space-time projectors
as~$E_1 = \text{diag}(1,1,0,0)$, $E_2 = \text{diag}(0,0,1,1)$ and consider a one-particle
fermionic operator~$P=-|\psi \ket \!\bra \psi|$. Then the systems obtained by
choosing~$\psi = (0,1,0,0)$ and~$\psi = (0,1,1,1)$ are not gauge equivalent, although they
give rise to the same particle representation.}. In this case, the two systems have identical causal
structures and give rise to exactly
the same densities and correlation functions. In other words, the two systems are indistinguishable by
any measurements, and thus one can take the point of view that they are equivalent descriptions
of the same physical system. Moreover, the particle representation gives a cleaner framework,
without the need for technical assumptions as in Definition~\ref{deftechnical}.
For these reasons, it seems preferable to take the point of view that the
particle representation is more fundamental, and to always deduce the space-time representation
by the constructions given after Definition~\ref{defparticle}.

\subsubsection{The Setting of the Fermionic Projector}
A particularly appealing special case is the setting of the {\em{fermionic projector}}, which we now
review\footnote{{\textsf{Note added on 1/27/2014}:}
Here by ``fermionic projector'' we mean that~$P$ is a projection operator in the Krein space~$(\K, \bra .|, \ket)$.
This projection property is built into the causal action principle
as the so-called identity constraint (see~\cite{continuum, lagrange}). However, as became clear
in the more recent papers~\cite{finite, infinite}, this projection property does not seem to be the
correct physical requirement. Instead, one should work with the mass normalization or
the spatial normalization of the fermionic projector. We refer to~\cite{norm} for details.}.
Beginning in the particle representation, we impose the additional constraint
\beq \label{prel}
\int_M x\, d\rho(x) = \1_\H\:,
\eeq
where the integral is assumed to converge in the strong sense, meaning that
\[ \int_M \|x\, \psi\|\: d\rho(x) < \infty \quad \text{for all $\psi \in \H$} \]
(where~$\| \psi \| = \sqrt{\la \psi | \psi \ra_\H}$ is the norm on~$\H$).
Under these assumptions, it is straightforward to verify from~\eqref{ndef} that the mapping
\[ \iota \::\: \H \rightarrow \K \:,\qquad (\iota \psi)(x) = \pi_x \psi \]
is well-defined. Moreover, the calculation
\[ \bra \iota \psi | \iota \psi \ket = \int_M \Sl \pi_x \psi \,|\, \pi_x \phi \Sr_x\: d\rho(x) 
\overset{\eqref{ssp}}{=} -\int_M \la \psi \,|\, x \phi \ra_\H \: d\rho(x) 
\overset{\eqref{prel}}{=} - \la \psi | \phi \ra_\H \]
shows that~$\iota$ is, up to a minus sign, an isometric embedding of~$\H$ into~$\K$.
Thus we may identify~$\H$ with the subspace~$\iota(\H) \subset \K$, and on this closed
subspace the inner products~$\Sl .|. \Sr_\H$ and~$\bra .|. \ket |_{\H \times \H}$ coincide up to a sign.
Moreover, the calculation
\[ (P \iota \psi)(x) = \int_M \pi_x \,y\, \pi_y \,\psi\, d\rho(y) = \int_M \pi_x \,y\, \psi\, d\rho(y)
= \pi_x \psi = (\iota \psi)(x) \]
yields that~$P$ restricted to~$\H$ is the identity. Next, for every~$\psi \in \D(P)$, the estimate
\begin{align*}
\Big\| \int_M y\, \psi(y)\, d\rho(y) \Big\|^2 & \overset{\eqref{prel}}{=}
\int_M d\rho(x) \iint_{M \times M} d\rho(y) \: d\rho(z)\;
\la y\, \psi(y) \,|\, x z\, \psi(z) \ra_\H \\
&\;\:=\; - \bra P \psi \,|\, P \psi \ket < \infty
\end{align*}
shows (after a straightforward approximation argument) that
\[ \phi := \int_M y\, \psi(y)\, d\rho(y) \;\in\; \H\:. \]
On the other hand, we know from~\eqref{Pdef} and~\eqref{Pxydef} that~$P \psi = \iota \phi$.
This shows that the image of~$P$ is contained in~$\H$. We conclude that~$P$ is a {\em{projection
operator}} in~$\K$ onto the negative definite, closed subspace~$\H \subset \K$.

\subsection{An Action Principle} \label{secaction}
We now return to the general setting of Definitions~\ref{defparticle} and~\ref{def2}.
For two points~$x, y \in \F$ we define the {\em{spectral weight}} $|.|$
of the operator products~$xy$ and~$(xy)^2$ by
\[ |xy| = \sum_{i=1}^{2n} |\lambda^{xy}_i|
\qquad \text{and} \qquad \left| (xy)^2 \right| = \sum_{i=1}^{2n} | \lambda^{xy}_i |^2 \:. \]
We also introduce the
\beq \label{Lagrange}
\text{\em{Lagrangian}} \qquad \L(x,y) = |(xy)^2| - \frac{1}{2n}\: |xy|^2 \:.
\eeq
For a given universal measure~$\rho$ on~$\F$, we define the non-negative functionals
\begin{align}
\text{\em{action}} \qquad \Sact[\rho] &= \iint_{\F \times \F} \L(x,y)\: d\rho(x)\, d\rho(y) \label{Sdef} \\
\text{\em{constraint}} \qquad \Tact[\rho] &= \iint_{\F \times \F} |xy|^2\: d\rho(x)\, d\rho(y) \label{Tdef}\:.
\end{align}
Our action principle is to
\beq \label{action}
\text{minimize~$\Sact$ for fixed~$\Tact$}\:,
\eeq
under variations of the universal measure.
These variations should keep the total volume unchanged, which means that a
variation~$(\rho(\tau))_{\tau \in (-\varepsilon, \varepsilon)}$
should for all~$\tau, \tau' \in (-\varepsilon, \varepsilon)$ satisfy the conditions
\[ \big| \rho(\tau) - \rho(\tau') \big|(\F) < \infty \qquad \text{and} \qquad
\big( \rho(\tau) - \rho(\tau') \big) (\F) = 0 \]
(where~$|.|$ denotes the total variation of a measure; see~\cite[\S28]{halmosmt}).
Depending on the application, one may impose additional constraints.
For example, in the setting of the fermionic projector, the variations should
obey the condition~\eqref{prel}. Moreover, one may prescribe properties of the universal measure
by choosing a measure space~$(\hat{M}, \hat{\mu})$ and restricting attention to universal measures
which can be represented as the push-forward of~$\hat{\mu}$,
\beq \label{pushforward}
\rho = F_* \hat{\mu} \qquad \text{with} \qquad \text{$F \::\: \hat{M} \rightarrow \F$ measurable}\:.
\eeq
One then minimizes the action under variations of the mapping~$F$.

The Lagrangian~\eqref{Lagrange} is compatible with our notion of causality in the
following sense. Suppose that two points~$x, y \in \F$ are spacelike separated
(see Definition~\ref{def2}). Then the eigenvalues~$\lambda^{xy}_i$ all have the same absolute value,
so that the Lagrangian~\eqref{Lagrange} vanishes. Thus pairs of points with spacelike separation do not
enter the action. This can be seen in analogy to the usual notion of causality where
points with spacelike separation cannot influence each other.

\subsection{Special Cases} \label{secspecial}
We now discuss modifications and special cases of the above setting as considered earlier.
First of all, in all previous papers except for~\cite{lqg} it was assumed that the Hilbert space~$\H$
is finite-dimensional and that the measure~$\rho$ is finite. Then by rescaling, one can
normalize~$\rho$ such that~$\rho(M)=1$. Moreover, the Hilbert space~$(\H, \la.|. \ra_\H)$ can
be replaced by~$\C^f$ with the canonical scalar product (the parameter~$f \in \N$ has the interpretation
as the number of particles of the system). These two simplifications lead to the
setting of {\em{causal variational principles}} introduced in~\cite{continuum}. More precisely, the
particle and space-time representations are considered in~\cite[Section~1 and~2]{continuum}
and~\cite[Section~3]{continuum}, respectively. The connection between the two representations
is established in~\cite[Section~3]{continuum} by considering the relation~\eqref{Fdef} in a
matrix representation.
In this context, $F$ is referred to as the {\em{local correlation matrix}} at~$x$.
Moreover, in~\cite{continuum} the universal measure is mainly represented as in~\eqref{pushforward}
as the push-forward of a mapping~$F$. This procedure is of advantage when analyzing the variational
principle, because by varying~$F$ while keeping~$(\hat{M}, \hat{\mu})$ fixed,
one can prescribe properties of the
measure~$\rho$. For example, if~$\hat{\mu}$ is a counting measure, then one varies~$\rho$ in the restricted
class of measures whose support consists of at most~$\# \hat{M}$ points with integer weights.
More generally, if~$\hat{\mu}$ is a discrete measure, then~$\rho$ is also discrete.
However, if~$\hat{\mu}$ is a continuous measure (or in more technical terms a so-called
non-atomic measure), then we do not get any constraints for~$\rho$, so that varying~$F$
is equivalent to varying~$\rho$ in the class of positive normalized regular Borel measures.

Another setting is to begin in the space-time representation (see Definition~\ref{def14}),
but assuming that~$\rho$ is a finite counting measure. Then the
relations~\eqref{Erel} become
\[ E_x E_y = \delta_{xy} E_x \qquad \text{and} \qquad \sum_{x \in M} E_x = \1_\K\:, \]
whereas the ``localization'' discussed after~\eqref{Erel}
reduces to multiplication with the space-time projectors,
\[ \psi(x) = E_x \psi\:, \qquad P(x,y) = E_x P E_y \:,\qquad \Sl \psi(x) \,|\, \phi(x) \Sr_x = \bra \psi \,|\, E_x\,
\phi \ket . \]
This is the setting of {\em{fermion systems in discrete space-time}} as considered in~\cite{PFP}
and~\cite{discrete, osymm}. We point out that all the work before 2006 deals with the
space-time representation, simply because the particle representation had not yet been found.

We finally remark that, in contrast to the settings considered previously, here the dimension and signature
of the spin space~$S_x$ may depend on~$x$. We only know that it is finite dimensional,
and that its positive and negative signatures are at most~$n$. In order to get into the setting of
constant spin dimension, one can isometrically embed every~$S_x$ into an indefinite inner
product space of signature~$(n,n)$ (for details see~\cite[Section~3.3]{continuum}).

\section{Spontaneous Structure Formation}
For a given measure~$\rho$, the structures of~$\F$ induce corresponding structures on
space-time~$M = \text{supp}\, \rho \subset \F$. Two of these structures are obvious:
First, on~$M$ we have the relative {\em{topology}} inherited from~$\F$.
Second, the causal structure on~$\F$ (see Definition~\ref{def2}) also induces a
{\em{causal structure}} on~$M$. Additional structures like a spin connection and
curvature are less evident; their construction will be outlined in Section~\ref{seclorentz} below.

The appearance of the above structures in space-time can also be understood as an effect of
{\em{spontaneous structure formation}} driven by our action principle.
We now explain this effect
in the particle representation (for a discussion in the space-time representation see~\cite{dice2010}).
For clarity, we consider the situation~\eqref{pushforward} where the universal
measure is represented as the push-forward of a given measure~$\hat{\mu}$ on~$\hat{M}$ (this
is no loss of generality because choosing a non-atomic measure space~$(\hat{M}, \hat{\mu})$,
any measure~$\rho$ on~$\F$ can be represented in this way;
see~\cite[Lemma~1.4]{continuum}).
Thus our starting point is a measure space~$(\hat{M}, \hat{\mu})$, without any additional structures.
The symmetries are described by the group of mappings~$T$ of the form
\beq \label{Tstruct}
T \::\: \hat{M} \rightarrow \hat{M} \quad \text{is bijective and preserves the measure~$\hat{\mu}$}\:.
\eeq
We now consider measurable mappings~$F : \hat{M} \rightarrow \F$ and minimize~$\Sact$ under
variations of~$F$. The resulting minimizer gives rise to a measure~$\rho = F_* \hat{\mu}$ on~$\F$.
On~$M:= \text{supp}\, \rho$ we then have the above structures inherited from~$\F$.
Taking the pull-back by~$F$, we get corresponding structures on~$\hat{M}$.
The symmetry group reduces to the mappings~$T$ which in addition to~\eqref{Tstruct}
preserve these structures. In this way, minimizing our action principle triggers an effect
of spontaneous symmetry breaking, leading to additional structures in space-time.

\section{A Lorentzian Quantum Geometry} \label{seclorentz}
We now outline constructions from~\cite{lqg} which give general notions
of a connection and curvature (see Theorem~\ref{thmspinconnection}, Definition~\ref{defmconn}
and Definition~\ref{defcurvature}). We also explain how these notions correspond to
the usual objects of differential geometry in Minkowski space (Theorem~\ref{thmminkowski})
and on a globally hyperbolic Lorentzian manifold (Theorem~\ref{thmglobhyp}).

\subsection{Construction of the Spin Connection}
Having Dirac spinors in a four-di\-men\-sio\-nal space-time in mind, we consider
as in Section~\ref{secmot} a causal fermion system of spin dimension two.
Moreover, we only consider space-time points $x \in M$ which are {\em{regular}}
in the sense that the corresponding spin spaces~$S_x$ have the maximal dimension four.

An important structure from spin geometry missing so far is Clifford multiplication.
To this end, we need a Clifford algebra represented by symmetric operators on~$S_x$.
For convenience, we first consider Clifford algebras with the maximal number of five
generators; later we reduce to four space-time dimensions (see Definition~\ref{defreduce} below).
We denote the set of symmetric linear endomorphisms of~$S_x$ by~$\Symm(S_x)$; it is
a $16$-dimensional real vector space.

\begin{Def}  {\em{ A five-dimensional subspace~$K \subset \Symm(S_x)$ is called
a {\em{Clifford subspace}} if the following conditions hold:
\begin{itemize}
\item[(i)] For any~$u, v \in K$, the anti-commutator~$\{ u,v \} \equiv u v + v u$ is a multiple
of the identity on~$S_x$.
\item[(ii)] The bilinear form~$\la .,. \ra$ on~$K$ defined by
\beq \label{anticommute2}
\frac{1}{2} \left\{ u,v \right\} = \la u,v \ra \, \1 \qquad {\text{for all~$u,v \in K$}}
\eeq
is non-degenerate and has signature~$(1,4)$.
\end{itemize} }}
\end{Def} \noindent
In view of the situation in spin geometry, we would like to distinguish a specific
Clifford subspace. In order to partially fix the freedom in choosing Clifford subspaces,
it is useful to impose that~$K$ should contain a given so-called sign operator.

\begin{Def} \label{defsign}  {\em{
An operator~$v \in \Symm(S_x)$ is called a {\em{sign operator}} if~$v^2 = \1$
and if the inner product~$\Sl .|v \,. \Sr\::\: S_x \times S_x \rightarrow \C$ is positive definite. }}
\end{Def}

\begin{Def}  {\em{
For a given sign operator~$v$, the set of {\em{Clifford extensions}} $\T^v$
is defined as the set of all Clifford subspaces containing~$v$,
\[ \T^v = \{K {\text{ Clifford subspace with }} v \in K \}\:. \]
}} \end{Def} \noindent
Considering~$x$ as an operator on~$S_x$, this operator has by definition of the spin
dimension two positive and two negative eigenvalues. Moreover, the calculation
\[ \Sl u | (-x) \,u \Sr_x \overset{\eqref{ssp}}{=} \la u | x^2 u \ra_\H > 0 \quad
\text{for all~$u \in S_x \setminus \{0\}$} \]
shows that the operator~$(-x)$ is positive definite on~$S_x$.
Thus we can introduce a unique sign operator~$s_x$ by demanding that
the eigenspaces of~$s_x$ corresponding to the
eigenvalues~$\pm 1$ are precisely the positive and negative spectral subspaces of
the operator~$(-x)$. This sign operator is referred to as the {\em{Euclidean sign operator}}.

A straightforward calculation shows that for two Clifford extensions $K, \tilde{K} \in \T^v$,
there is a unitary transformation~$U \in e^{i \R v}$ such that~$\tilde{K} = U K U^{-1}$
(for details see~\cite[Section~3]{lqg}). By dividing out this group action, we obtain a five-dimensional
vector space, endowed with the inner product~$\la ., \ra$. Taking for~$v$ the Euclidean signature
operator, we regard this vector space as a generalization of the usual tangent space.
\begin{Def} \label{deftangent}  {\em{
The {\em{tangent space}}~$T_x$ is defined by
\[ T_x = \T_x^{s_x} / \exp(i \R s_x)\:. \]
It is endowed with an inner product~$\la .,. \ra$ of signature~$(1,4)$. }}
\end{Def}

We next consider two space-time points, for which we need to make the following assumption.
\begin{Def} \label{defproptl}  {\em{
Two points~$x,y \in M$ are said to be {\em{properly time-like}}
separated if the closed chain~$A_{xy}$ has a strictly positive spectrum and if the corresponding
eigenspaces are definite subspaces of~$S_x$. }}
\end{Def} \noindent
This definition clearly implies that~$x$ and~$y$ are time-like separated (see Definition~\ref{def2}).
Moreover, the eigenspaces of~$A_{xy}$ are definite if and only if those of~$A_{yx}$ are,
showing that Definition~\ref{defproptl} is again symmetric in~$x$ and~$y$.
As a consequence, the spin space can be decomposed uniquely
into an orthogonal direct sum~$S_x = I^+ \oplus I^-$ of a positive definite subspace~$I^+$ and a
negative definite subspace~$I^-$ of~$A_{xy}$.
This allows us to introduce a unique sign operator~$v_{xy}$ by demanding
that its eigenspaces corresponding to the eigenvalues~$\pm 1$
are the subspaces~$I^\pm$. This sign operator is referred to as the
{\em{directional sign operator}} of~$A_{xy}$.
Having two sign operators~$s_x$ and~$v_{xy}$ at our disposal, we can distinguish unique
corresponding Clifford extensions, provided that the two sign operators satisfy the following
generic condition.
\begin{Def} \label{defgensep}  {\em{
Two sign operators~$v, \tilde{v}$ are said to be {\em{generically separated}} if
their commutator~$[v, \tilde{v}]$ has rank four. }}
\end{Def}

\begin{Lemma} \label{lemma3}
Assume that the sign operators~$s_x$ and~$v_{xy}$ are generically separated.
Then there are unique Clifford extensions~$K_x^{(y)} \in \T^{s_x}$ and~$K_{xy} \in \T^{v_{xy}}$ 
and a unique operator~$\rho \in K_x^{(y)} \cap K_{xy}$ with the following properties:
\begin{itemize}
\item[(i)] The relations~$\{ s_x, \rho \} = 0 = \{ v_{xy}, \rho \}$ hold. \\[-1em]
\item[(ii)] The operator~$U_{xy} := e^{i \rho}$ transforms one Clifford extension to the other,
\beq \label{tKKrho}
K_{xy} = U_{xy} \,K_x^{(y)}\, U_{xy}^{-1}\:.
\eeq
\item[(iii)] If~$\{s_x, v_{xy}\}$ is a multiple of the identity,
then~$\rho=0$.
\end{itemize}
The operator~$\rho$ depends continuously on~$s_x$ and~$v_{xy}$.
\end{Lemma} \noindent
We refer to~$U_{xy}$ as the {\em{synchronization map}}.
Exchanging the roles of~$x$ and~$y$, we also have two sign operators~$s_y$ and~$v_{yx}$
at the point~$y$. Assuming that these sign operators are again generically separated,
we also obtain a unique Clifford extension~$K_{yx} \in \T^{v_{yx}}$.

After these preparations, we can now explain the construction of the spin connection~$D$
(for details see~\cite[Section~3]{lqg}).
For two space-time points~$x,y \in M$ with the above properties, we want to introduce
an operator
\beq \label{csc}
D_{x,y} \::\: S_y \rightarrow S_x
\eeq
(generally speaking, by the subscript~$_{xy}$ we always denote an object at the point~$x$, whereas
the additional comma $_{x,y}$ denotes an operator which maps an object at~$y$
to an object at~$x$).
It is natural to demand that~$D_{x,y}$ is unitary, that~$D_{y,x}$ is its inverse, and that
these operators map the directional sign operators at~$x$ and~$y$ to each other,
\begin{align}
D_{x,y} &= (D_{y,x})^* = (D_{y,x})^{-1} \label{Dc1} \\
v_{xy} &= D_{x,y}\, v_{yx}\, D_{y,x}\:. \label{Dc2}
\end{align}
The obvious idea for constructing an operator with these properties is to take a polar
decomposition of~$P(x,y)$; this amounts to setting
\beq \label{Dfirst}
D_{x,y} = A_{xy}^{-\frac{1}{2}}\: P(x,y)\:.
\eeq
This definition has the shortcoming that it is not compatible with the chosen Clifford extensions.
In particular, it does not give rise to a connection on the corresponding tangent spaces. In order to
resolve this problem, we modify~\eqref{Dfirst} by the ansatz
\beq \label{Dsecond}
D_{x,y} = e^{i \varphi_{xy}\, v_{xy}}\: A_{xy}^{-\frac{1}{2}}\: P(x,y)
\eeq
with a free real parameter~$\varphi_{xy}$. In order to comply with~\eqref{Dc1}, we need to
demand that
\beq \label{phicond}
\varphi_{xy} = -\varphi_{yx} \!\!\!\mod 2 \pi \:;
\eeq
then~\eqref{Dc2} is again satisfied. We can now use the freedom in choosing~$\varphi_{xy}$
to arrange that the distinguished Clifford subspaces~$K_{xy}$ and~$K_{yx}$ are mapped onto
each other,
\beq \label{Dc3}
K_{xy} = D_{x,y} \:K_{yx}\: D_{y,x}\:.
\eeq
It turns out that this condition determines~$\varphi_{xy}$ up to multiples of~$\frac{\pi}{2}$.
In order to fix~$\varphi_{xy}$ uniquely in agreement with~\eqref{phicond}, we need 
to assume that~$\varphi_{xy}$ is not a multiple of~$\frac{\pi}{4}$. This leads us to the
following definition.
\begin{Def} \label{defspinconnect}
{\em{ Two points~$x,y \in M$ are called {\em{spin connectable}} if
the following conditions hold:
\begin{itemize}
\item[(a)] The points~$x$ and~$y$ are properly timelike separated
(note that this already implies that~$x$ and~$y$ are regular
as defined at the beginning of Section~\ref{seclorentz}).
\item[(b)] The Euclidean sign operators~$s_x$ and~$s_y$ are generically separated
from the directional sign operators~$v_{xy}$ and~$v_{yx}$, respectively.
\item[(c)] Employing the ansatz~\eqref{Dsecond}, the phases~$\varphi_{xy}$
which satisfy condition~\eqref{Dc3} are not multiples of~$\frac{\pi}{4}$.
\end{itemize} 
}}
\end{Def} \noindent
We denote the set of points which are spin connectable to~$x$ by~$\I(x)$.
It is straightforward to verify that~$\I(x)$ is an open subset of~$M$.

Under these assumptions, we can fix~$\varphi_{xy}$ uniquely by imposing that
\beq \label{Dc4}
\varphi_{xy} \in \Big(-\frac{\pi}{2}, -\frac{\pi}{4} \Big) \cup \Big(\frac{\pi}{4}, \frac{\pi}{2} \Big)\:,
\eeq
giving the following result (for the proofs see~\cite[Section~3.3]{lqg}).

\begin{Thm} \label{thmspinconnection}
Assume that two points~$x,y \in M$ are spin connectable. Then there is a unique
{\bf{spin connection}}~$D_{x,y} : S_y \rightarrow S_x$ of the form~\eqref{Dsecond}
having the properties~\eqref{Dc1}, \eqref{Dc2}, \eqref{Dc3} and~\eqref{Dc4}.
\end{Thm}

\subsection{A Time Direction, the Metric Connection and Curvature}
We now outline a few further constructions from~\cite[Section~3]{lqg}.
First, for spin connectable points we can distinguish a direction of time.
\begin{Def} \label{deftime} {\em{ Assume that the points~$x, y \in M$ are spin connectable.
We say that~$y$ lies in the {\em{future}} of~$x$ if the phase~$\varphi_{xy}$
as defined by~\eqref{Dsecond} and~\eqref{Dc4} is positive.
Otherwise, $y$ is said to lie in the {\em{past}} of~$x$. }}
\end{Def} \noindent
According to~\eqref{phicond}, $y$ lies in the future of~$x$ if and only if~$x$ lies in the
past of~$y$. By distinguishing a direction of time, we get a structure similar to
a causal set (see for example~\cite{sorkin}). However, in contrast to a causal set, our notion of
``lies in the future of'' is not necessarily transitive.

The spin connection induces a connection on the corresponding tangent spaces, as we now explain.
Suppose that~$u_y \in T_y$. Then, according to Definition~\ref{deftangent} and Lemma~\ref{lemma3},
we can consider~$u_y$ as a vector of the representative~$K_y^{(x)} \in \T^{s_y}$.
By applying the synchronization map, we obtain a vector in~$K_{yx}$,
\[ u_{yx} := U_{yx} \,u_y\, U_{yx}^{-1} \in K_{yx}\:. \]
According to~\eqref{Dc3}, we can now ``parallel transport'' the vector to the Clifford
subspace~$K_{xy}$,
\[ u_{xy} := D_{x,y} \, u_{yx}\, D_{y,x} \in K_{xy}\:. \]
Finally, we apply the inverse of the synchronization map to obtain the vector
\[ u_x := U_{xy}^{-1} \,u_{xy}\, U_{xy} \in K_x^{(y)} \:. \]
As~$K_x^{(y)}$ is a representative of the tangent space~$T_x$ and all transformations were unitary,
we obtain an isometry from~$T_y$ to~$T_x$. 
\begin{Def} \label{defmconn} {\em{ The isometry between the tangent spaces defined by
\[ \nabla_{x,y} \::\: T_y \rightarrow T_x \::\: u_y \mapsto u_x \]
is  referred to as the {\em{metric connection}} corresponding to the spin connection~$D$. }}
\end{Def} \noindent

We next introduce a notion of curvature.
\begin{Def} \label{defcurvature}  {\em{
Suppose that three points~$x, y, z \in M$ are pairwise spin connectable. Then the
associated {\em{metric curvature}} $R$ is defined by
\beq \label{Rmdef}
R(x,y,z) = \nabla_{x,y} \,\nabla_{y,z} \,\nabla_{z,x} \::\: T_x \rightarrow T_x\:.
\eeq
}} \end{Def} \noindent
The metric curvature~$R(x,y,z)$ can be thought of as a discrete analog of the
holonomy of the Levi-Civita connection on a manifold, where a tangent vector is parallel
transported along a loop starting and ending at~$x$. On a manifold, the curvature at~$x$ is immediately
obtained from the holonomy by considering the loops in a small neighborhood of~$x$.
With this in mind, Definition~\ref{defcurvature} indeed generalizes the usual notion of curvature to
causal fermion systems.

The following construction relates directional sign operators to vectors of the tangent space.
Suppose that $y$ is spin connectable to~$x$. By synchronizing the directional sign operator~$v_{xy}$,
we obtain the vector
\beq \label{yxdef}
\hat{y}_x := U_{xy}^{-1} \,v_{xy}\, U_{xy} \in K_x^{(y)} \:.
\eeq
As~$K_x^{(y)} \in \T^{s_x}$ is a representative of the tangent space, we can regard~$\hat{y}_x$ as a
tangent vector. We thus obtain a mapping
\[ \I(x) \rightarrow T_x \;:\; y \mapsto \hat{y}_x \:. \]
We refer to~$\hat{y}_x$ as the {\em{directional tangent vector}} of~$y$ in~$T_x$.
As~$v_{xy}$ is a sign operator and the transformations in~\eqref{yxdef} are unitary,
the directional tangent vector is a timelike unit vector with the additional property that
the inner product~$\Sl .| \hat{y}_x . \Sr_x$ is positive definite.

We finally explain how to reduce the dimension of the tangent space to four, with the
desired Lorentzian signature~$(1,3)$. 
\begin{Def} {\em{ The fermion system is called {\em{chirally symmetric}} if to every~$x \in M$
we can associate a spacelike vector~$u(x) \in T_x$ which is orthogonal to all directional tangent vectors,
\[ \la u(x), \hat{y}_x \ra = 0 \qquad \text{for all~$y \in \I(x)$} \:, \]
and is parallel with respect to the metric connection, i.e.
\[ u(x) = \nabla_{x,y} \,u(y)\, \nabla_{y,x} \qquad \text{for all~$y \in \I(x)$} \:. \] }}
\end{Def}

\begin{Def} \label{defreduce} {\em{
For a chirally symmetric fermion system, we introduce the {\em{reduced tangent
space}} $T_x^\text{red}$ by
\[ T_x^\text{red} = \langle u_x \rangle^\perp \subset T_x \:. \] }}
\end{Def} \noindent
Clearly, the reduced tangent space has dimension four and signature~$(1,3)$.
Moreover, the operator~$\nabla_{x,y}$ maps the reduced tangent spaces isometrically to each other.
The local operator~$\gamma^5 := -i u/\sqrt{-u^2}$ takes the role of the {\em{pseudoscalar matrix}}.

\subsection{The Correspondence to Lorentzian Geometry} \label{secCLG}
We now explain how the above spin connection is related to the
usual spin connection used in spin geometry (see for example~\cite{lawson+michelsohn, baum}).
To this end, let~$(M, g)$ be
a time-oriented Lorentzian spin manifold with spinor bundle~$SM$
(thus~$S_xM$ is a $4$-dimensional complex vector space endowed with an
inner product~$\Sl .|. \Sr_x$ of signature~$(2,2)$). Assume that~$\gamma(t)$
is a smooth, future-directed and timelike curve, for simplicity parametrized by the arc length, defined
on the interval~$[0, T]$ with~$\gamma(0) = y$ and~$\gamma(T) = x$.
Then the parallel transport of tangent vectors along~$\gamma$ with respect to the
Levi-Civita connection~$\nablaLC$ gives rise to the isometry
\[ \nablaLC_{x,y} \::\: T_y \rightarrow T_x \:. \]
In order to compare with the metric connection~$\nabla$ of Definition~\ref{defmconn},
we subdivide~$\gamma$ (for simplicity with equal spacing, although a non-uniform spacing
would work just as well). Thus for any given~$N$, we define the
points $x_0, \ldots, x_N$ by
\[ x_n = \gamma(t_n) \qquad \text{with} \qquad t_n = \frac{n T}{N}\:. \]
We define the parallel transport~$\nabla_{x,y}^N$ by successively composing
the parallel transport between neighboring points,
\[ \nabla^N_{x,y} := \nabla_{x_N, x_{N-1}} \nabla_{x_{N-1}, x_{N-2}} \cdots \nabla_{x_1, x_0}
\::\: T_y \rightarrow T_x \:. \]

Our first theorem gives a connection to the Minkowski vacuum. For any~$\varepsilon>0$
we regularize on the scale~$\varepsilon>0$ by inserting a convergence generating factor
into the integrand in~\eqref{Psea},
\beq \label{Pseaeps}
P^\varepsilon(x,y) = \int \frac{d^4k}{(2 \pi)^4}\: (k \slsh+m)\:
\delta(k^2-m^2)\: \Theta(-k^0)\: e^{\varepsilon k^0}\: e^{-ik(x-y)} \:.
\eeq
This function can indeed be realized as the kernel of the fermionic operator~\eqref{Pxydef}
corresponding to a causal fermion system~$(\H, \F, \rho^\varepsilon)$.
Here the measure~$\rho^\varepsilon$ is the push-forward of the volume measure
in Minkowski space by an operator~$F^\varepsilon$, being an ultraviolet regularization
of the operator~$F$ in~\eqref{bdef}-\eqref{Fmapdef} (for details see~\cite[Section~4]{lqg}).

\begin{Thm} \label{thmminkowski}
For given~$\gamma$, we consider the family of
regularized fermionic projectors of the vacuum~$(P^\varepsilon)_{\varepsilon>0}$
as given by~\eqref{Pseaeps}.
Then for a generic curve~$\gamma$ and for every~$N \in \N$, there is~$\varepsilon_0$
such that for all~$\varepsilon \in (0, \varepsilon_0]$ and all~$n=1,\ldots, N$, the
points~$x_{n}$ and~$x_{n-1}$ are spin connectable,
and~$x_{n+1}$ lies in the future of~$x_n$ (according to Definition~\ref{deftime}).
Moreover,
\[ \nablaLC_{x,y} = \lim_{N \rightarrow \infty} \:\lim_{\varepsilon \searrow 0} \nabla^N_{x,y} \:. \]
\end{Thm} \noindent
By a {\em{generic curve}} we mean that the admissible curves are dense in the
$C^\infty$-topology (i.e., for any smooth~$\gamma$ and every~$K \in \N$,
there is a sequence~$\gamma_\ell$ of admissible curves such that~$D^k \gamma_\ell \rightarrow D^k \gamma$
uniformly for all~$k =0, \ldots, K$). The restriction to generic curves is needed in order to
ensure that the Euclidean and directional sign operators are generically separated
(see Definition~\ref{defspinconnect}~(b)).
The proof of the above theorem is given in~\cite[Section~4]{lqg}.

Clearly, in this theorem the connection~$\nablaLC_{x,y}$ is trivial. In order to
show that our connection also coincides with the Levi-Civita connection
in the case with curvature, in~\cite[Section~5]{lqg} a globally hyperbolic Lorentzian manifold
is considered. For technical simplicity, we assume that the manifold is flat Minkowski space
in the past of a given Cauchy hypersurface.

\begin{Thm} \label{thmglobhyp} Let $(M,g)$ be a globally hyperbolic manifold which is
isometric to Minkowski space in the past of a given Cauchy-hypersurface~${\mathcal{N}}$.
For given~$\gamma$, we consider the family of
regularized fermionic projectors~$(P^\varepsilon)_{\varepsilon>0}$ such that~$P^\varepsilon(x,y)$
coincides with the distribution~\eqref{Pseaeps} if~$x$ and~$y$ lie in the past of~${\mathcal{N}}$.
Then for a generic curve~$\gamma$
and for every sufficiently large~$N$, there is~$\varepsilon_0$
such that for all~$\varepsilon \in (0, \varepsilon_0]$ and all~$n=1,\ldots, N$, the
points~$x_{n}$ and~$x_{n-1}$ are spin connectable, and~$x_{n+1}$ lies in the future
of~$x_n$ (according to Definition~\ref{deftime}). Moreover,
\beq \label{eqLorentz}
\lim_{N \rightarrow \infty} \:\lim_{\varepsilon \searrow 0} \nabla^N_{x,y} - \nablaLC_{x,y} 
= \O \!\left( L(\gamma)\: \frac{\nabla R}{m^2} \right)
\Big(1 + \O \Big( \frac{\text{\rm{scal}}}{m^2} \Big) \Big) \:,
\eeq
where~$R$ denotes the Riemann curvature tensor, ${\text{\rm{scal}}}$ is scalar curvature,
and~$L(\gamma)$ is the length of the curve~$\gamma$.
\end{Thm} \noindent
Thus the metric connection of Definition~\ref{defmconn} indeed coincides with the
Levi-Civita connection, up to higher order curvature corrections.
For detailed explanations and the proof we refer to~\cite[Section~5]{lqg}.

We conclude this section by pointing to a few additional constructions in~\cite{lqg} which cannot be
explained consistently in this short survey article. First, there is the subtle point that the unitary
transformation~$U \in \exp(i \R s_x)$ which is used to identify two representatives~$K, \tilde{K}
\in T_x$ via the relation~$\tilde{K} = U K U^{-1}$ (see Definition~\ref{deftangent})
is not unique. More precisely, the operator~$U$ can be transformed according to
\[ U \rightarrow -U \qquad \text{and} \qquad U \rightarrow s_x \,U\:. \]
As a consequence, the metric connection (see Definition~\ref{defmconn}) is defined only up
to the transformation
\[ \nabla_{x,y} u \rightarrow s_x \,(\nabla_{x,y} u)\, s_x\:. \]
Note that this transformation maps representatives of the same tangent vector into each other,
so that~$\nabla_{x,y} u \in T_x$ is still a well-defined tangent vector.
But we get an ambiguity when composing the metric connection several times
(as for example in the expression for the metric curvature in Definition~\ref{defcurvature}).
This ambiguity can be removed by considering {\em{parity-preserving systems}}
as introduced in~\cite[Section~3.4]{lqg}.

At first sight, one might conjecture that Theorem~\ref{thmglobhyp} should also apply to
the spin connection in the sense that
\beq \label{conjecture}
\DLC_{x,y} = \lim_{N \rightarrow \infty} \:\lim_{\varepsilon \searrow 0} D^N_{x,y} \:,
\eeq
where~$\DLC$ is the spin connection on~$SM$ induced by the Levi-Civita connection and
\beq \label{spinit}
D^N_{x,y} := D_{x_N, x_{N-1}} D_{x_{N-1}, x_{N-2}} \cdots D_{x_1, x_0}
\::\: S_y \rightarrow S_x
\eeq
(and~$D$ is the spin connection of Theorem~\ref{thmspinconnection}).
It turns out that this conjecture is false. But the conjecture becomes true if we
replace~\eqref{spinit} by the operator product
\[ D^N_{(x,y)} := D_{x_N, x_{N-1}} U_{x_{N-1}}^{(x_N | x_{N-2})}
D_{x_{N-1}, x_{N-2}} U_{x_{N-2}}^{(x_{N-1} | x_{N-3})}
\cdots U_{x_1}^{(x_2 | x_0)} D_{x_1, x_0} \:. \]
Here the intermediate factors~$U_{.}^{(.|.)}$ are the so-called {\em{splice maps}}
given by
\[ U_x^{(z|y)} = U_{xz} \,V\, U_{xy}^{-1} \:, \]
where~$U_{xz}$ and~$U_{xy}$ are synchronization maps, and~$V \in \exp(i \R s_x)$
is an operator which identifies the representatives~$K_{xy}, K_{xz} \in T_x$
(for details see~\cite[Section~3.7 and Section~5]{lqg}).
The splice maps also enter the {\em{spin curvature}}~$\mathfrak{R}$, which is defined in
analogy to the metric curvature~\eqref{Rmdef} by
\[ \mathfrak{R}(x,y,z) = U_x^{(z|y)} \:D_{x,y}\: U_y^{(x|z)} \:D_{y,z}\: U_z^{(y|x)}
\:D_{z,x} \::\: S_x \rightarrow S_x \:. \]

\section{A ``Quantization Effect'' for the Support of Minimizers} \label{secsupport}
\begin{figure}
 \includegraphics[width=8cm]{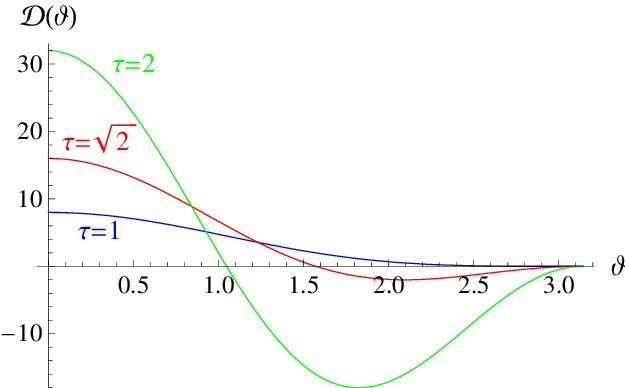}
 \caption{The function $\D$.}
 \label{plotD}
\end{figure}

The recent paper~\cite{support} contains a first numerical and analytical study of the minimizers
of the action principle~\eqref{action}. We now explain a few results and discuss their potential
physical significance. We return to the setting of causal variational principles (see
Section~\ref{secspecial}). In order to simplify the problem as much as possible, we only consider
the case of spin dimension~$n=1$ and two particles~$f=2$ (although many results in~\cite{support}
apply similarly to a general finite number of particles). Thus we identify the particle
space~$(\H, \la .|. \ra_\H)$ with~$\C^2$. Every point~$F \in \F$ is a Hermitian $(2 \times 2)$-matrix
with at most one positive and at most one negative eigenvalue. We represent it in terms
of the Pauli matrices as~$F = \alpha \, \1 + \vec{u} \vec{\sigma}$ with~$|\vec{u}| \geq |\alpha|$.
In order to further simplify the problem, we prescribe the eigenvalues in the support of the universal
measure to be~$1+\tau$ and~$1-\tau$, where~$\tau \geq 1$ is a given parameter. Then~$F$
can be represented as
\[ F = \tau\: x \!\cdot\! \sigma + \1 \qquad \text{with} \qquad x \in S^2 \subset \R^3 \:. \]
Thus we may identify~$\F$ with the unit sphere~$S^2$.
The Lagrangian~$\L(x,y)$ given in~\eqref{Lagrange} simplifies to a function
only of the angle~$\vartheta$ between the vectors~$x$ and~$y$. More precisely,
\begin{align*}
\L(x,y) &= \max \big( 0, \D(\langle x,y \rangle) \big) \\
\D(\cos \vartheta) &= 2 \tau^2\: (1+ \cos \vartheta) \left( 2 - \tau^2 \:(1 - \cos \vartheta) \right) .
\end{align*}
As shown in Figure~\ref{plotD} in typical examples,
the function~$\D$ is positive for small~$\vartheta$
and becomes negative if~$\vartheta$ exceeds a certain value~$\vartheta_{\max}(\tau)$.
Following Definition~\ref{def2}, two points~$x$ and~$y$ are timelike
separated if~$\vartheta<\vartheta_{\max}$ and spacelike separated if~$\vartheta > \vartheta_{\max}$.

\begin{figure}
\includegraphics[width=10cm]{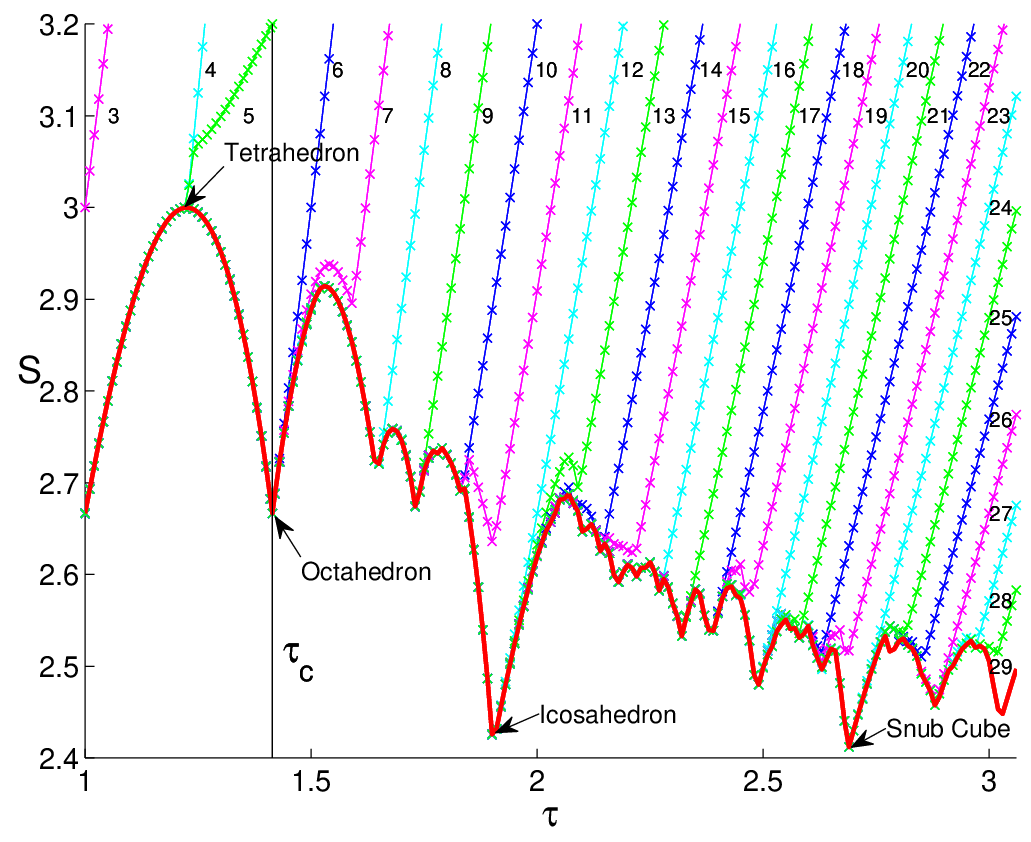}
\caption{Numerical minima of the action on the two-dimensional sphere.}
\label{figvgl_weight}
\end{figure}

Our action principle is to minimize the action~\eqref{Sdef} by varying the measure $\rho$ in the family of normalized Borel measures on the sphere.
In order to solve this problem numerically, we approximate the minimizing measure by
a weighted counting measure. Thus for any given integer~$m$, we choose points~$x_1, \ldots, x_m
\in S^2$ together with corresponding weights~$\rho_1, \ldots, \rho_m$ with
\[ \rho_i \geq 0 \qquad \text{and} \qquad \sum_{i=1}^m \rho_i = 1 \]
and introduce the measure~$\rho$ by
\beq \label{weightcount}
\rho = \sum_{i=1}^m \rho_i\: \delta_{x_i} \:,
\eeq
where~$\delta_x$ denotes the Dirac measure.
Fixing different values of~$m$ and seeking for numerical minimizers by varying
both the points~$x_i$ and the weights~$\rho_i$,
we obtain the plots shown in Figure~\ref{figvgl_weight}.
It is surprising that for each fixed~$\tau$, the obtained minimal action
no longer changes if~$m$ is increased beyond a certain value~$m_0(\tau)$. The numerics shows
that if~$m>m_0$, some of the points~$x_i$ coincide, so that the support of the minimizing
measure never consists of more than~$m_0$ points. Since this property remains true in the
limit~$m \rightarrow \infty$, our numerical analysis shows that the
{\em{minimizing measure is discrete}} in the sense that its support consists of a finite
number of~$m_0$ points. Another interesting effect is that the action seems to favor
symmetric configurations on the sphere. Namely, the most distinct
local minima in Figure~\ref{figvgl_weight} correspond to configurations where
the points~$x_i$ lie on the vertices of Platonic solids.
The analysis in~\cite{support} gives an explanation for this ``discreteness'' of the minimizing
measure, as is made precise in the following theorem (for more general results in the same spirit
see~\cite[Theorems~4.15 and~4.17]{support}).
\begin{Thm} If~$\tau > \tau_c := \sqrt{2}$, the support of
every minimizing measure on the two-dimensional sphere is singular in the sense that
it has empty interior.
\end{Thm} \noindent

Extrapolating to the general situation, this result indicates
that our variational principle favors discrete over continuous configurations.
Again interpreting~$M := \text{supp}\, \rho$ as our space-time, this might correspond
to a mechanism driven by our action principle which makes space-time discrete.
Using a more graphic language, one could say that space-time ``discretizes itself''
on the Planck scale, thus avoiding the ultraviolet divergences of quantum field theory.

Another possible interpretation gives a connection to field quantization:
Our model on the two-sphere involves one continuous parameter~$\tau$. If we allow~$\tau$
to be varied while minimizing the action (as is indeed possible if we drop the constraint
of prescribed eigenvalues), then the local minima of the action attained at discrete values of~$\tau$
(like the configurations of the Platonic solids) are favored. Regarding~$\tau$ as the
amplitude of a ``classical field'', our action
principle gives rise to a ``quantization'' of this field, in the sense that the amplitude
only takes discrete values.

The observed ``discreteness'' might also account for effects related to the wave-particle duality
and the collapse of the wave function in the measurement process (for details see~\cite{dice2010}).

\section{The Correspondence to Quantum Field Theory and Gauge Theories} \label{secQFT}
The correspondence to Minkowski space mentioned in Section~\ref{secCLG}
can also be used to analyze our action principle for interacting systems in the so-called
{\em{continuum limit}}. We now outline a few ideas and constructions
(for details see~\cite[Chapter~4]{PFP}, \cite{sector} and the survey article~\cite{srev}).
We first observe that the vacuum fermionic projector~\eqref{Psea}
is a solution of the Dirac equation~$(i \gamma^j \partial_j - m) P^\text{sea}(x,y)=0$.
To introduce the interaction, we replace the free Dirac operator by a more general Dirac operator,
which may in particular involve gauge potentials or a gravitational field.
For simplicity, we here only consider an electromagnetic potential~$A$,
\beq \label{DiracP}
\left( i \gamma^j (\partial_j - ie A_j) - m \right) P(x,y) = 0 \:.
\eeq
Next, we introduce particles and anti-particles by occupying (suitably normalized)
positive-energy states and removing states of the sea,
\beq \label{particles}
P(x,y) = P^\text{sea}(x,y)
-\frac{1}{2 \pi} \sum_{k=1}^{n_f} |\psi_k(x) \Sr \Sl \psi_k(y) |
+\frac{1}{2 \pi} \sum_{l=1}^{n_a} |\phi_l(x) \Sr \Sl \phi_l(y) | \:.
\eeq
Using the so-called causal perturbation expansion and light-cone expansion,
the fermio\-nic projector can be introduced uniquely from~\eqref{DiracP} and~\eqref{particles}.

It is important that our setting so far does not involve the field equations; in particular,
the electromagnetic potential in the Dirac equation~\eqref{DiracP} does not need to satisfy
the Maxwell equations. Instead, the field equations should be derived from
our action principle~\eqref{action}. Indeed, analyzing the corresponding Euler-Lagrange equations,
one finds that they are satisfied only if the potentials in the Dirac equation satisfy certain
constraints. Some of these constraints are partial differential equations involving
the potentials as well as the wave functions of the particles and anti-particles in~\eqref{particles}.
In~\cite{sector}, such field equations are analyzed in detail for a system involving an
axial field. In order to keep the setting as simple as possible, we here consider the analogous
field equation for the electromagnetic field
\beq \label{Maxwell}
\partial_{jk} A^k - \Box A_j = e \sum_{k=1}^{n_f} \Sl \psi_k | \gamma_j \psi_k \Sr
-e \sum_{l=1}^{n_a} \Sl \phi_l | \gamma_j \phi_l \Sr \:.
\eeq
With~\eqref{DiracP} and~\eqref{Maxwell}, the interaction as described by the
action principle~\eqref{action} reduces in the continuum limit to the
{\em{coupled Dirac-Maxwell equations}}.
The many-fermion state is again described by the fermionic projector, which is built up
of {\em{one-particle wave functions}}. The electromagnetic field merely is a
{\em{classical bosonic field}}. Nevertheless, regarding~\eqref{DiracP} and~\eqref{Maxwell}
as a nonlinear hyperbolic system of partial differential equations and treating
it perturbatively, one obtains all the Feynman diagrams which do not involve fermion loops.
Taking into account that by exciting sea states we can describe pair creation and annihilation
processes, we also get all diagrams involving fermion loops.
In this way, we obtain agreement with perturbative quantum field theory
(for details see~\cite[\S8.4]{sector} and the references therein).

We finally remark that in the continuum limit, the freedom in choosing the spinor basis~\eqref{lgf}
can be described in the language of standard gauge theories. Namely, introducing a gauge-covariant
derivative~$D_j = \partial_j - i C_j$ with gauge potentials~$C_j$
(see for example~\cite{pokorski}), the transformation~\eqref{lgt}
gives rise to the local gauge transformations
\beq \begin{split}
\psi(x) &\rightarrow U(x)\, \psi(x) \:,\qquad\quad D_j \rightarrow U D_j U^{-1} \\
C_j(x) &\rightarrow U(x) C(x) U(x)^{-1} + i U(x)\, (\partial_j U(x)^{-1})
\end{split} \label{gauge}
\eeq
with~$U(x) \in \U(p,q)$. The difference to standard gauge theories is that the gauge group cannot
be chosen arbitrarily, but it is determined to be the isometry group of the spin space.
In the case of spin dimension two, the corresponding gauge group~$\U(2,2)$ allows for a unified
description of electrodynamics and general relativity (see~\cite[Section~5.1]{PFP}). By
choosing a higher spin dimension (see~\cite[Section~5.1]{PFP}), one gets a larger gauge group.
Our mathematical framework ensures that our action principle and thus also the continuum limit
is {\em{gauge symmetric}} in the sense that the transformations~\eqref{gauge} with~$U(x) \in \U(p,q)$
map solutions of the equations of the continuum limit to each other.
However, our action is {\em{not}} invariant under local transformations of the form~\eqref{gauge}
if~$U(x) \not \in \U(p,q)$ is not unitary.  An important example of such non-unitary transformations
are chiral gauge transformations like
\[ U(x) = \chi_L\, U_L(x) + \chi_R\, U_R(x) \quad \text{with} \quad
U_{L\!/\!R} \in \U(1) \:, \;\; U_L \not \equiv U_R\:. \]
Thus chiral gauge transformations do not describe a gauge symmetry in the above sense.
In the continuum limit, this leads to a mechanism which gives chiral gauge fields a rest mass
(see~\cite[Section~8.5]{sector} and~\cite[Section~7]{srev}). Moreover, in systems of higher
spin dimension, the presence of chiral gauge fields gives rise to a spontaneous breaking of the
gauge symmetry, resulting in a smaller ``effective'' gauge group. As shown in~\cite[Chapters~6-8]{PFP},
these mechanisms make it possible to realize the gauge groups and couplings of the standard model.

\Thanks{{\em{Acknowledgments:}}  We thank the referee for helpful suggestions on the manuscript.}

\providecommand{\bysame}{\leavevmode\hbox to3em{\hrulefill}\thinspace}
\providecommand{\MR}{\relax\ifhmode\unskip\space\fi MR }
\providecommand{\MRhref}[2]{%
  \href{http://www.ams.org/mathscinet-getitem?mr=#1}{#2}
}
\providecommand{\href}[2]{#2}

\end{document}